\documentclass[11pt]{article}
\usepackage{graphics,epsfig}

\font\num=msbm10

\begin{document}


\title{The Liar-paradox in a Quantum Mechanical Perspective\footnote{Appeared in Foundations of
Science, 4(2), 1999, 115-132.}}
\author{Diederik Aerts\footnote{Senior Research Associate of the Fund for
Scientific Research, Flanders}, Jan
Broekaert\footnote{The research for this paper was realized with the aid of
AWI-grant Caw96/54a of the Flemish
Community. }, Sonja Smets
\footnote{Research Assistant of the Fund for Scientific Research Flanders}}
\date {}
\maketitle
\centerline{CLEA, Free University of  Brussels, Pleinlaan 2, B-1050 Brussels}
\centerline{e-mails: diraerts@vub.ac.be, jbroekae@vub.ac.be,
sonsmets@vub.ac.be}

\begin{abstract}
\noindent In this paper we concentrate on the nature of the liar paradox as
a cognitive entity; a consistently
testable configuration of properties.  We elaborate further on a quantum
mechanical model [Aerts, Broekaert,
Smets 1999] that has been proposed to analyze the dynamics involved, and we
focus on the interpretation and
concomitant philosophical picture. Some conclusions we draw from our model
favor an effective realistic
interpretation of cognitive reality.

\end{abstract}

\section{Introduction.}
\par\noindent  Our approach, to analyze the Liar paradox as a cognitive
entity, emerges naturally from
research on integrating world views (Worldviews Group, 1994, 1995;
Broekaert, 1999). The justification
is based on their necessary inclusion in an encompassing model.   An
integrating world view can be based on a
model of interacting layers epistemically corresponding to the various
contemporary sciences (CLEA,1997).
Basically, in the same way as we view social entities in a social layer, or
quantum entities in the quantum
--- which in this context can be viewed as the pre-material --- layer,  we
place cognitive entities in a
cognitive layer of reality. Which is considered the expanse where the
personal and interpersonal cognitive
interactions are taking place, and which will be elaborated in the next
section.
\par\noindent In this paper, we will work with versions of the Liar
Sentence with index or sentence pointers
followed by the sentence this index points at. Typically we will either
concentrate on the single sentence
version;

\begin{center} {\sl `Single Liar'}

\medskip

(1)  \ \ \    sentence (1) is false

\end{center} or the two sentence version;
\begin{center}
\bigskip {\sl `Double Liar'}

\medskip
 (1)   \ \ \    sentence (2) is false

\medskip
 (2)    \ \ \   sentence (1) is true

\end{center}
\par\noindent  First, we stress that it is not our intention to construct a
solution for the paradox, but we
will concentrate in this paper on a model of the self-referential
circularity --- more precisely, the
truth-value dynamics --- involved.  In this way, we are able  to
understand the nature of the liars'
paradoxicality by  modelling the typical truth-value oscillation in the
encompassing framework.  By
taking this position there is no need to go into details of
`solving'-strategies that have been proposed in
the literature. Actually, the most important of these strategies are
specifically constructed to avoid the
paradox by, for example, declaring it `ungrammatical', or `meaningless' or
`ungrounded'. All these
strategies, of which we only mentioned some, and those that introduced
truth value `gaps' or `gluts', have
proven valuable, and serve here to distinguish the present position. For
more details on the mentioned solving
strategies, see for example  (Grim 1991, chp 1).
\par\noindent Finally, we mention some issues concerning the recourse
taken to the quantummechanical
formalism in describing cognitive entities. Contrary to the present paper,
many attempts have been made to
approach `emergent' phenomena --- like cognition --- through complex
dynamics. Focusing on the collective
dynamics of quite different entities, some very similar patterns are
encountered regardless of the nature of
the substrate. This has been pointed out in research pertaining to the
theories of  nonlinearity   and
complexity.  Because of the efficacy of these theories, tentative
conclusions about the nature of  --- quite
disparate --- processes in reality can be inferred from
structural-dynamical similarities.  Although, we have
to stress that these similarities generally involve only so called
`noncontextual'  entities, as the
theories of  nonlinearity and complexity  are  fundamentally classical
theories. By
noncontextuality  we mean that the entity will `not' be influenced by an
act of observation (measurement of
the entity). A direct consequence of the fact that the theories of
nonlinearity  and complexity are classical
deterministic.  It is now our claim that cognitive entities --- especially
of the type of the liar paradox
--- share with quantum entities a variable, and sometimes high, contextual
nature:  i.e. also cognitive
entities, as it is the case for quantum entities, are influenced by the act
of measurement, which in the case
of a cognitive entity is generated by the cognitive interaction.  Therefore
the theory of
quantum mechanics can be used in analyzing the nature of cognitive entities
and more specifically the
nature of the liar paradox.  In (Aerts, Broekaert, Smets, 1999) we built a
quantum mechanical model that
fits this purpose.  With respect to the high contextual nature of intricate
cognitive and also social
entities, we can point out an approach where the probability model that
results in an opinion pole  --- with
important influence of the interviewer on the interviewee --- is of a
quantum mechanical nature (Aerts,1998;
Aerts and Aerts,1995,1996; Aerts, Coecke and Smets,1999).  In a similar
way, we encounter an important
contextual influence of an observer reading (observing) the liar paradox
sentences.  In this case the high
contextuality will occur through the observer --- the cognitive person ---
reasoning through the
self-referent entity.

\section{Cognitive Entities}

The existence of a cognitive entity is recognized by its aptitude of being
generally and cognitively
influenced on as a practically stable configuration, e.g. in the reasoning
on it and communicating about it,
and by the limited number of different states that it can be in. The
cognitive entity is endowed with
properties and relations with the other elements of its layer. Their
internal coherence relates to language
as well as conceptions of experience. The extent to which it is related to
pendants in the physical and other
layers enhances its identity and its granted coincidence with the
physically real entity. This variable
correspondence relation is put forward between entities of the ontological
cognitive layer and their pendants
in physical or other layers.

\par\noindent In order to give a more precise and complete characterization
of the cognitive entity, we need
to reconstruct its interaction profile. This is done by specifying the
different kinds of measurements or
experiments appropriate to the intended entity. This necessarily happens
with the intervention of the
cognitive person --- the conscious human being conceiving the cognitive entity.

\par\noindent We distinguish between those entities that refer to elements
of layers other than the cognitive
one and entities that do not. Arguably, such a distinction may only be
possible in specific phases of
cognition and theory forming, respecting personal psychological evolution.
In both cases, the influencing or
intervening of the cognitive person will be different; engendering physical
interventions or only cognition,
reasoning.
\par\noindent We define such an intervention as a kind of measurement on
the entity with the aim of gaining
information or increasing knowledge on its precise state. In order to keep
our analogy with the physical
layer, we restrict our exposition here to measurements that characterize
the state of a cognitive entity.

\par\noindent The state formalizes specific --- or all ---  possible
actualisations according the interaction
profile. The exhaustive description being practically impossible, an
idealized state description can
only be obtained for the overall state. We mention however that this
problem appears in an analogous way
in physics, where in each model also only an idealized state description is
obtained, depending on the
possible experiments that are available.   Hence, a specific property state
corresponds, as in physics, to
appropriate measurements that are available.    In the case of the liar
paradox entity, the exact meaning
of the truth-state will be explained in the next section.   The
measurements that characterize the state of
an entity referring to the classical-material layer are mostly
straightforward.  For example, analyzing if a
piece of chalk is breakable has a fixed measurement-procedure namely,
breaking the piece of chalk.   This
gives the corresponding entity a classical behavior in the sense that each
time we perform the measurement we
obtain, with certainty, the same result.   In a quantum mechanical context
we say that such entities are
in  an `eigenstate' for the corresponding measurement. Also cognitive
entities that do not refer to other
layers can have straightforward measurement-procedures.  Especially in the
case of e.g. a mathematical
theorem,  there is often a straightforward procedure to establish a proof
for it. And indeed each time
we apply the same procedure, we obtain the same result, so we say again
that those entities are in
 an eigenstate corresponding to this measurement procedure. Naturally, we
do not expect everybody to
know these mathematics, so the state of that same theorem will depend on
what we call the `cognitive
background' of each cognitive person. Even mathematicians will have
different opinions on the status of some
theorems, therefore it is clear that the general cognitive background of
people, over history, is important
and plays a role in the nature of the states that will be attributed to
cognitive entities. As this example
points out, the nonfixed character of  knowledge implies we accord states
of nonfixed character to cognitive
entities. Again however, we remark  that this is analogous to the physical
situation. The state of a
physical entity, defined correspondingly to measurement procedures, depends
also on the general body of
experimental possibilities at a certain epoch of time in the history of
human culture, and hence is not fixed
once and forever. Depending on whether this state has been defined by more
`universal' methods of
experimentation, the state will approach in a deeper way the ontology of
the entity. The same holds true for
the state of a cognitive entity.

\par\noindent  In the case of the liar paradox as a cognitive entity, for
any measurement procedure, the
truth-value will give us different results each time we intervene with it.
Here, the state describing the
truth-value will be called a superposition state as related to the
measurement procedure in question, in
analogy with the quantum mechanical concept. This means that we can not
obtain a `certain' prediction --- in
a classical sense --- of properties, in this case `thruth' of `false', of
that cognitive entity. The input of
the quantummechanical formalism is therefore appropriate.

\par\noindent How does one measure the liar paradox? The measurement here
consists of two part-processes,
'reading the sentence' and `making a sentence true or false'.  This means
that in our description the liar
paradox within the cognitive layer of reality is `in general' --- before
the measurement --- not in a
state such that a reading would give true or false. The `true state ' and
the `false state ' of the
sentence are specific states; `eigenstates' of the measurement. In general,
the state of the liar
paradox is not one of these two eigenstates.   Due to the act of
measurement, and in analogy with what
happens during a quantum measurement, the state of the sentence changes
(`collapses') into one of the two
possible eigenstates, the `true state' or the `false state'. This act of
making a sentence true or false
can be specifically described as `read it and make an hypothesis about its
truth or falsehood'. In  the next
section we will apply this approach to the Double Liar, and see that an
initial measurement followed by the
sequence of logical inferences puts into work an oscillation dynamics that
we can describe by a
Schr{\"o}dinger evolution over reasoning-time. We will also see in the next
section, that the change of
state due to measurement can be described by a projection operator in the
quantum mechanical Hilbert space
where the Schr{\"o}dinger evolution is defined.

\par\noindent  Once we make an hypothesis about one sentence the whole
entity starts changing from one
truth-state into another by continued reading with logical inference. When
we stop this process, by means of
not `looking at' or `reasoning on' the sentences any more, the entity ---
we hypothesise --- reestablishes
its original superposition state of indefiniteness.   This superposition
state does not correspond to what
has been called a truth value gap and does not assign a third truth value
to the liar. In this sense, in
our model, the set of semantical truth values, of which only one can  be
assigned to the cognitive entity,
does not contain a third truth value or a value gap.

\par\noindent Finally, we remark some interpretative issues concerning the
origin of the entity's dynamics
and the nature of the cognitive layer.  From the quantummechanical analogy,
the temporal evolution of the
entity is expected to originate intrinsically, still the  construction of
the evolution --- as will be clear
from the next section --- supposes the cognitive person's motivation by
reasoning. We interpret the latter to
be reflected in the autonomous dynamics, as such, the cognitive entity
obtains its cognitive essence. The
precise origin of temporal evolution has thereby become less transparent;
the entity as well as the cognitive
person will engender identical evolution.
\par\noindent The nature of the cognitive layer, is essentially different
from the sphere spanned by common
material objects. In our approach --- akin to `effective' realism --- the
cognitive layer is  `Hilbert-space
like', a personal and mental construct with social and cultural
conditioning, a collective dynamic emergent
layer carried by its cognitive participators. The extent and subtlety of
this issue, allows in the present
context merely explanatory simplifications. More detailed elaborations of
the cognitive layer as an emergent,
Hilbert-space like sphere in social groups are due (Aerts, Broekaert and
Gabora, 1999).

\section{The Liar-paradox: A Quantum Description of its Truth Behavior.}

\par\noindent We will first discuss the Single Liar entity. By `measuring'
the single sentence we attribute a
chosen truth-value to the sentence, immediately and logically inferring
from its lecture the opposite
truth-value. The cognitive person is therefore inclined to attribute in an
alternating manner opposite truth
values to the Single Liar sentence, until the `measuring' process is chosen
to be stopped. Subsequently no
decisive and unambiguous truth-value can be attributed to the Single Liar
sentence as it is. The application
of the quantummechanical formalism suggests to describe this situation by a
superposition of opposite
truth-value states. We remark the striking correspondence between  truth
values and the two-fold eigenvalues
of a spin-1/2 state of some quantum particles (e.g. an electron), and the
oscillatory dynamics present in the
reasoning dynamics respectively the evolution dynamics of a
 spin-1/2 particle in a constant magnetic field.  This formal
correspondence will be used to construct a
dynamical representation of the cognitive entity. Recall that this
correspondence is possible, due to the
fact that the spin of a particle is a quantized property, it exposes itself
by means of distinct spin values.
For a spin-1/2 particle there are only two distinct spin values namely `up'
and `down' --- appropriately
corresponding to `true' and `false' values for the cognitive entity. The
quantum mechanical formalism enables
to express a superposition of these `up' and `down' states of a spin-1/2
particle by simple addition. When we
apply the same idea to the single liar sentence, we obtain a state
$\Psi$ described by a pondered superposition of the two states of opposite
truth-value:

\[
 \Psi = c_{ true} \left( \begin{array}{c}  1 \\ 0 \end{array}
\right)  +
 c_{ false} \left(\begin{array}{c}  0 \\ 1 \end{array} \right)
\]
\par\noindent The measurement itself, namely the interaction of the
cognitive person on the entity when the
sentence is being made true or being made false is described respectively
by the true-projector
$P_{true}$ or false-projector
$P_{false}$.

\[
 P_{true} = \left( \begin{array}{cc}  1 & 0 \\ 0 & 0
  \end{array} \right)  \ \ \ \ \ \  P_{false } = \left(
\begin{array}{cc}  0 & 0 \\ 0 & 1
  \end{array} \right)
\]
\par\noindent In this quantum mechanical description, the true-measurement
(false measurement) on the
superposed state
$\Psi$ results in the true state (resp. false state).  The true state is
represented as follows :
\[
 P_{true} \Psi = c_{true}\left( \begin{array}{c}  1 \\ 0
\end{array} \right)
\]
\par\noindent where the square modulus of the corresponding pondering
factor $c_{ true}$ gives the
statistical probability of finding the entity in the true-state. An
unequivocal result is therefore not
obtained when the superposition does not leave out one of the states
completely, i.e. either
$c_{true}$ or $c_{false}$  is zero. Only in those instances do we
unambiguously attribute to a sentence its
truth or falsehood.\\
\par\noindent We now look at the Double Liar sentences in more formal and
mathematical detail. We consider
three situations:
\[ {\rm A}\ \   \left\{
\begin{array} {ll}
 {\rm (1)\  }   &  {\rm sentence\ (2)\ is\ false} \\ {\rm (2)\ }   &   {\rm
sentence\ (1)\ is\ true}
\end{array} \right.
\]

\[ {\rm B}\ \   \left\{
\begin{array} {ll}
 {\rm (1)\  }   &  {\rm sentence\ (2)\ is\ true} \\ {\rm (2)\ }   &  {\rm
sentence\ (1)\ is\ true}
\end{array} \right.
\]

\[ {\rm C} \ \   \left\{
\begin{array} {ll}
 {\rm (1)\  }   &  {\rm sentence\ (2)\ is\ false} \\ {\rm (2)\ }   &   {\rm
sentence\ (1)\ is\ false}
\end{array} \right.
\] From the case of the Single Liar we expect here a representation of the
truth-behaviour by coupled {\num
C}$^2$ vectors, one for each sentence.  Closer inspection of the coupled
sentences of the two-sentence liar
paradox of type (B) and (C), shows this is possible. The measurement of the
Double Liar (B) allways will
couple true-states of (B1) and (B2),  and false-states of (B1) and (B2). In
the case (C) on the other hand,
measurement will couple  the true-state of (B1) to the false-state of (B2),
and the false-state of (B1) to
the true-state of (B2). From a formal point of view, the equivalent
spin-states in quantum mechanics would be
described by the so called `singlet state' and a `triplet state'
respectively. The singlet state indicates
that two spin-1/2 particles are anti-alined and in an anti-symmetrical
state , while a triplet state
indicates that two spin-1/2 particles are  alined and in a symmetrical state.
\par\noindent Whereas the Single Liar has been mathematically represented
in a {\num C}$^2$ finite dimensional
complex Hilbert Space, we now need a {\num C}$^2 \otimes$\,{\num C}$^2$
space for the description of the (B)
or (C) Double Liar. The tensorproduct $\otimes$ connects the two sentences
into one composed entity.
\par\noindent In the specific case of (C), and taking into account the
anti-symmetric spin analog,
$\Psi$ is written as:
\[
\frac{1}{\sqrt{2}}\left\{ \left( \begin{array}{c}  1  \\ 0
\end{array} \right)  \otimes \left(
\begin{array}{c}  0 \\ 1 \end{array} \right)  -  \left(
\begin{array}{c}  0  \\ 1 \end{array} \right)
\otimes \left( \begin{array}{c}  1 \\ 0 \end{array} \right)  \right\}
\] Still, in the application to the cognitive entity other choices of the
pondering coefficients are
possible. The only constraints on the coefficients are: equal amplitude and
addition of the squared
amplitudes to unity.
\par\noindent The state-vector for the liar paradox in case (B) can be
constructed in a similar manner:
\[
\frac{1}{\sqrt{2}}\left\{ \left( \begin{array}{c}  1  \\ 0
\end{array} \right)  \otimes \left(
\begin{array}{c}  1 \\ 0
\end{array} \right)  +  \left( \begin{array}{c}  0  \\ 1 \end{array}
\right)  \otimes \left(
\begin{array}{c}  0 \\ 1 \end{array} \right)  \right\}
\]

\par\noindent The projection operators which make sentence one and
respectively sentence two true are now:
\[
 P_{1,true} = \left( \begin{array}{cc}  1 & 0 \\ 0 & 0
  \end{array} \right) \otimes {\bf 1}_2 \ \ \ \ \ \  P_{2,true } = {\bf 1}_1
\otimes \left( \begin{array}{cc}  1 & 0
\\ 0 & 0
  \end{array} \right)
\]
\par\noindent The projection operators that make the sentences false are
obtained by switching the  elements
$1$ and $0$ on the diagonal of the matrix.  These four projection operators
represent the possible `logical'
interactions between the cognitive person and the cognitive entity.
\par\noindent During the continued measurement on the entities (B) and (C),
the sequence of logical
inferences results in a repetitive pattern of consecutive true-false
states.  These patterns are not very
complicated, in case of entity (B) it will be a repetition of true-states
(resp. false-states) depending on
whether we presupposed an initial true (resp. false) state.  While in the
case of entity (C) it will always
be an alternation between true-states and false states, no matter which
state we presupposed.

\par\noindent Finally, we describe in detail the original double liar
paradox, case (A). In this case we will
show how the true-false cycle originates from the Schr{\"o}dinger
time-evolution of the appropriate initial
state.
\par\noindent Instead of working within the coupled Hilbert space
{\num C}$^2 \otimes$\,{\num C}$^2$, as in cases (B) and (C), we have to use
a space of higher dimension for
(A). This complexification is due to the fact that no initial state can be
found in the restricted space
{\num C}$^2 \otimes$\,{\num C}$^2$, such that application of the four
true-false projection operators results
in four orthogonal states respectively representing the four
truth-falsehood states. The existence of such a
superposition state --- with equal amplitudes of its components ---  is
required to describe the entity prior
to, and after, any measurement procedure.  If we perform the continued
measurement on (A), by consecutive
logical inference,  the dynamical pattern is not anymore a two-step process
like in the previous cases (B)
and (C),instead we have a four-step process.  Starting from the initial
superposition state this four-step
process can not be described by the coupled spin-1/2 models any more. In
order to resolve this problem,
recourse has to be taken to a 4 dimensional Hilbert-space for each sentence.
The Hilbert-space needed to
describe the Double Liar (A) is therefore {\num C}$^4 \otimes$\,{\num C}$^4$.
\par\noindent The initial un-measured superposition state --- $\Psi_0$ ---
of the Double Liar (A) is given by
any equally pondered superposition of the four true-false states:

{\small\[
\frac{1}{2}\left\{ \left( \begin{array}{c} 0\\ 0\\ 1 \\ 0 \end{array}
\right)  \otimes \left(
\begin{array}{c}  0 \\ 1 \\ 0 \\ 0 \end{array} \right)  +
 \left( \begin{array}{c} 0\\ 1\\ 0 \\ 0 \end{array} \right)  \otimes
\left( \begin{array}{c}  0 \\ 0 \\ 0
\\ 1 \end{array} \right)  +
\left( \begin{array}{c} 0\\ 0\\ 0 \\ 1 \end{array} \right)  \otimes
\left( \begin{array}{c} 1 \\ 0 \\ 0
\\ 0
\end{array} \right) +
\left( \begin{array}{c} 1\\ 0\\ 0 \\ 0 \end{array} \right)  \otimes
\left( \begin{array}{c}  0 \\ 0 \\ 1
\\ 0
\end{array} \right) \right\}
\]}

\par\noindent Each next term in this superposition state is the consecutive
state which is reached in the
course of time, when the paradox is reasoned through. The truth-falsehood
values attributed to these states,
refer to the chosen measurement projectors.
\par\noindent Making a sentence true or false in the act of measurement,
will be described by the appropriate
projection operators in {\num C}$^4 \otimes$\,{\num C}$^4$. In the case we
make sentence 1 (resp. sentence 2)
true we get:
\[
 P_{1,true} = \left( \begin{array}{cccc}  0 & 0 & 0 & 0 \\ 0 & 0 & 0 & 0 \\
0 & 0 & 1 & 0 \\0 & 0 & 0 & 0
  \end{array} \right) \otimes {\bf 1}_2 \ \ \ \ \ \
 P_{2,true } = {\bf 1}_1 \otimes\left( \begin{array}{cccc}  0 & 0 & 0 & 0
\\ 0 & 0 & 0 & 0 \\ 0 & 0 & 1 & 0 \\ 0 & 0 & 0 & 0 \end{array} \right)
\] The projectors for the false-states are constructed by placing the
$1$ on the final diagonal place:
\[
 P_{1,false} = \left( \begin{array}{cccc}  0 & 0 & 0 & 0 \\ 0 & 0 & 0 & 0
\\ 0 & 0 & 0 & 0 \\0 & 0 & 0 & 1
  \end{array} \right) \otimes {\bf 1}_2 \ \ \ \ \ \
 P_{2,false} = {\bf 1}_1 \otimes\left( \begin{array}{cccc}  0 & 0 & 0 & 0
\\ 0 & 0 & 0 & 0 \\ 0 & 0 & 0 & 0 \\ 0 & 0 & 0 & 1 \end{array} \right)
\]
\par\noindent As a consequence of making a freely chosen sentence of (A)
either true or false, by logical
inference the four consecutive states are repeatedly run trough.  In order
to give a time-ordered description
of this cyclic change of state, a continous time $t$ is introduced as an
ordering parameter. The time-odering
parameter extrapolates the discrete moments of consecutive outcomes of the
logical inferences, and as such
relates to the physical time of reasoning. Under these interpretative
restrictions a Schr\"odinger evolution
over `time' can be constructed.
\par\noindent  Essentially, a Hamiltonian $H$ can be constructed, such that
the unitary evolution operator
$U(t)$ --- with
 $U(t) = e^{-iHt}$ --- describes the cyclic change of logical inferences.
\\ The construction of the
evolution operator $U(t)$ is more easily accomplished by switching
temporarily to an equivalent
representation in a  larger Hilbert space. Switching to a {\num C}$^{16}$
Hilbert-space is done without any
modification or alteration of the problem, as it is isomorphic to the
original {\num C}$^4 \otimes$\,{\num
C}$^4$ coupled Hilbert space.
\par\noindent A new basis in {\num C}$^{16}$ is constructed from the basis of
{\num C}$^4 \otimes$\,{\num C}$^4$ (  $i$ and $j$ from 1 to 4 )  :\[  e_i
\otimes e_j = e_{
\kappa(i ,j) } \ \ \ {\rm and} \ \ \ \kappa (i,j) =  4(i-1) +j
\]  Where $\kappa$ is the natural basis transformation function from the
$C^4\otimes C^4$ to the {\num C}$^{16}$ Hilbert space, and the index from
the new basis states in {\num
C}$^{16}$.
\\ The initial superposition state $\Psi_0$ can now be  represented in
{\num C}$^{16}$ by:
\[
\Psi_0 = \frac{1}{2} \{ e_{10} + e_{8} + e_{13} + e_{3} \}
\]
\par\noindent For notational ease we continue to work further in a
4-dimensional subspace of {\num C}$^{16}$,
namely this subspace generated by the basis $(e_{10}, e_{8}
,e_{13},e_{3})$. Obviously we do not loose any
information by this restriction. The 4 by 4 submatrix --- $U_D$ --- of the
discrete unitary evolution
operator, which describes the time-evolution at  discrete instants of time
when consecutive outcomes of
logical inferences have been reached, is:
\[  U_D =
\left( \begin{array}{cccc}   0 & 0 & 0 & 1 \\   0 & 0 & 1 & 0 \\ 1 & 0 & 0 & 0
\\ 0 & 1 & 0 & 0
\end{array}\right)
\]
\par\noindent  In order to obtain a description at every instance of the
time-ordering parameter, a procedure
of diagonalisation on the submatrix $U_D$ is performed, i.e. $ U_D |_{\rm
diag}$. The diagonalisation
procedure allows to solve the matrix equation by breaking it into four
uncoupled  scalar equations. From the
Schr\"odinger evolution and Stone's Theorem we obtain:
\[
 H_{sub} |_{\rm diag} =i \ln  U_D|_{\rm diag}
\]   Inverting the procedure of diagonalisation, the infinitesimal
generator of the time-evolution --- the
submatrix hamiltonian --- is obtained :
\[  H_{sub} =
\left( \begin{array}{cccc}    -1/2&-1/2&(1-i)/2&(1+i)/2 \\
-1/2&-1/2&(1+i)/2&(1-i)/2 \\
(1+i)/2&(1-i)/2&1/2&1/2
\\ (1-i)/2&(1+i)/2&1/2&1/2
\end{array}\right)
\]
\par\noindent Or in terms of gamma matrices (these are merely introduced
for shorthand notation for there is
no implication of relativistic nature):

\[  H_{sub} = \frac{1}{2} \left( - \gamma_0 - \gamma_5 +  \gamma_0\gamma_1
\right) +
\frac{i}{2} \left( \gamma_1+ \gamma_2\gamma_3 +  \gamma_0\gamma_5 \right)
\]

\par\noindent The submatrix of the evolution operator $U(t)$, valid at all
intermediary times $t$ is then
given by the expression:
\[  U_{sub}(t) = e^{- i H_{sub} t}
\]
\par\noindent The time evolution operator $U_{sub}(t)$ in the 4-dimensional
subspace of {\num C}$^{16}$ is:
\begin{eqnarray}  U_{sub}(t) & = & \frac{1}{4} \left\{
\left(1 + e^{-i t} + e^{i t} + e^{ 2 i t}\right) 1  + i \left(1 - e^{-i t}
- e^{i t} + e^{ 2 i t}\right)
\gamma_2\gamma_3 \right. \nonumber \\
 &  &\left. - \left(1 - e^{ 2 i t}\right) \gamma_5  + i  \left( e^{-i t} -
e^{i t}\right)\gamma_0 \gamma_5  -
i
\left( - e^{-i t} + e^{i t}\right)\gamma_1
\right\} \nonumber
\end{eqnarray}
\par\noindent In order to finalise our initial claim, we should bring back
the hamiltonian $H$ as well as the
time-evolution operator $U(t)$ in the original {\num C}$^4 \otimes$\,{\num
C}$^4$ Hilbert space. Although the
outcome is straightforward to obtain by using the  basis transformation
function $\kappa (i, j)$,  the
complexity of the expression necessitates shorthand notation:
\[
 H =\sum_{\kappa, \lambda = 1}^{16} {H _{sub}}_{\kappa(i,j) \lambda(u,v)}
O_{i u}\otimes O_{j v}
\] and
\[
 U(t) =\sum_{\kappa, \lambda = 1}^{16} {U _{sub}}_{\kappa(i,j)
\lambda(u,v)}(t) O_{i u}\otimes O_{j v}
\] with;
\[ O_{i u}\otimes O_{j v}=\{ e_i.e_u^t \}\otimes\{ e_j.e_v^t \}
\] For example, the term $\kappa = 3$ , $\lambda  = 10$ of the time
evolution operator $U(t)$ is given by;
\[
\frac{1}{4}(1- i e^{-i t} +i e^{-i t} - i e^{ 2 i t})
\left( \begin{array}{cccc}   0 & 0 & 1 & 0 \\   0 & 0 & 0 & 0 \\ 0 & 0 & 0 & 0
\\ 0 & 0 & 0 & 0
\end{array}\right) \otimes \left( \begin{array}{cccc}   0 & 0 & 0 & 0 \\ 0
& 0 & 0 & 0 \\ 0 & 1 & 0 & 0
\\ 0 & 0 & 0 & 0
\end{array}\right)
\]
\par\noindent

\par\noindent Finally, we complete the dynamical picture of the Double Liar
cognitive entity (A); when
submitted to any measurement at choice, the entity starts its
truth-falsehood cycle, when left un-measured
the entity remains statically in its undifferentiated superposition state.
The latter statement follows
immediately from the fact that the initial state
$\Psi_0$ is left unchanged by the dynamical evolution $U(t)$;
\[
\Psi_0 (t) = \Psi_0
\]
$\Psi_0$ is a time invariant, as it is an eigenstate of the Hamiltonian
$H$. Exactly this time invariance points to the fact that the state
$\Psi_0$ describes the existence of the
cognitive entity (A) in its cognitive space, independent of any observer.
The highly contextual nature of
the  Double Liar (A) --- its unavoidable dynamics engendered by measuring
it --- implies,  intrinsically it
can not expose its complete nature, analogous to the quantum entities of
the micro-physical world.

The evolution over the time-ordering parameter $t$ of the truth
behaviour of the cognitive entity during the measurement process can be
illustrated graphically.

\includegraphics{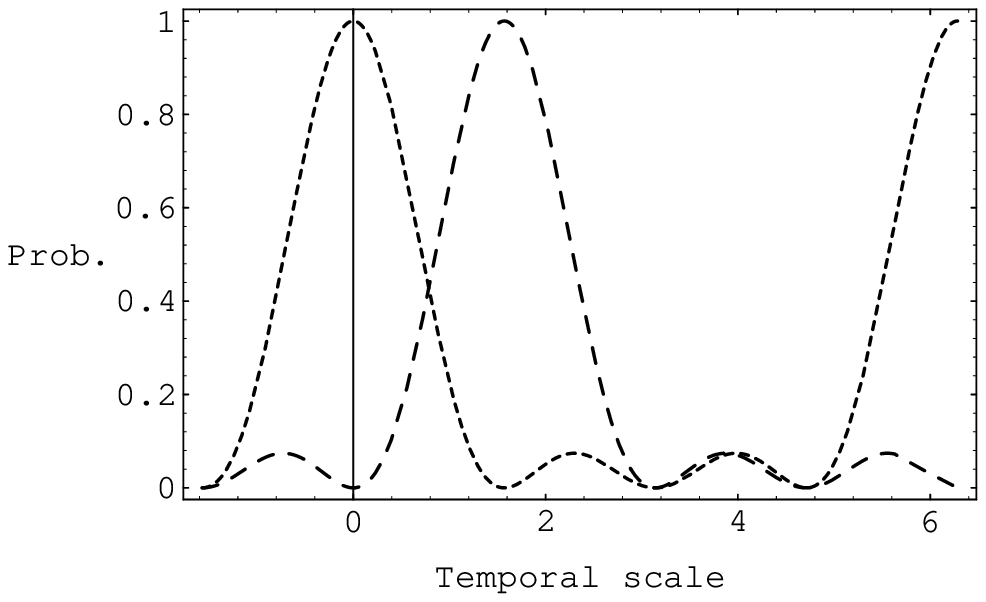}

\noindent {\small \em Temporal evolution
of the cognitive evaluation of the
Double Liar (A). At time-ordering parameter $t = 0$, a `false'-measurement
on (A1) has been executed on the
initial state $\Psi_0$ (short-dashed line). The second consecutive step in
the logical inference, the
`true'-state of (A2) at time-ordering parameter $t = \frac{Pi}{2}$, and the
intermediary times, is shown by
the long-dashed line. For clarity, consecutive steps have been deleted.
Discrete moments  of outcomes of
logical inferences are at time-ordering parameter
$t = n \frac{\pi}{2}$}

\section{Conclusion.}

We found that the quantum mechanical formalism can be applied  to
self-referent cognitive entities as the
single liar and double liar paradox. Essentially the two necessary features
of the dynamics were constructed;
when measured the entity starts its truth-falsehood cycle, when left
un-measured the entity remains
invariantly in its initial state. A measurement of the entity ---
engendered by appropriating one sentence
its truth or falsehood --- sets into action the dynamical evolution which
attributes, alternatively over
time, truth and falsehood to the coupled sentences. The unmeasured entity
$\Psi_0$ does not change over time,
its invariance reflects its emergent nature. \\

\par \noindent The truth-falsehood alternation, due to a measurement
process, and the time-invariance of the
initial state  are both derived from constructed evolution operator $U(t)$,
driven by the Hamiltonian $H$.
The quantum formalism therefore has proven an appropriate tool to describe
the liar paradox entity.   \\

\par \noindent We set out with the idea of realist cognitive entity, if it
can be generally and cognitively
influenced on as a practically stable configuration, and by the limited
number of different states that it
can be in. We put forward that exactly this time invariance points to the
fact that the state $\Psi_0$
describes the existence of the cognitive entity in its cognitive space,
independent of any observer. For this
reason we take the position at liberty to put forward a realistic picture
of the cognitive reality, in the
sense that $\Psi_0 (t)$ represents the state of the real entity.\\

\par \noindent The nature of the cognitive layer, as here proposed, is
essentially different from the space
of common macroscopic material objects. The cognitive layer is
`Hilbert-space like', and originates
from the cognitive person. Quite different from the complexity-theory
approach, which emphasizes dynamics, the
present formalism clearly describes the emergent  entity as an ontological
state. An important question in our
research therefore remains; the relation between the quantum mechanical
state description and the non-linear
approach of complex dynamics. Rather speculatively some relation between
privileged states in both models
could be expected. A model exposing common grounds to both formalisms could
in first approach allow inquiry
into the relation between the quantum mechanical eigenstates of the
Hamiltonian and the phase-space
attractors of the corresponding non-linear evolution equations. Eventually,
more detailed elaborations of the
cognitive layer as an emergent, Hilbert-space like sphere in social groups,
can relate both approaches.\\

\par \noindent   The generalisation to other cognitive entities with
reference to non-cognitive entities or
processes can most probably, to extent, be covered by the $\epsilon$-model
formalism (Aerts, 1986, Aerts and
Durt 1994, Aerts, Durt and Van Bogaert 1993).  This model allows to
describe effects of intermediary
contextuality, which is expected in common cognitive entities.  Further
development needs to bring into the
formalism, a general  dynamic,  and contextual influences other than those
of the cognitive person on the
truth behavior of the cognitive entities.\\

\par\noindent Philosophical questions, quite  speculative at this stage of
our research, can be put forward:
e.g. Can we, from the example of the Liar paradox, learn something in
general about the nature and origin of
dynamical change? What real life experiments can underpin the realistic
approach to the cognitive entities?

\section{References.}

Aerts, D., (1986),  ``A possible explanation for the probabilities of quantum
mechanics'', {\it J. Math. Phys.}, {\bf 27}, 202.

\smallskip
\noindent
Aerts, D. (1992), `` Construction of Reality and its Influence on the
Understanding of Quantum Structures'', {\it  Int. J.  Theor. Phys.}, Vol
31, 10, p.
1813.

\smallskip
\noindent
Aerts, D. (1994), ``The biomousa: a new view of discovery and
creation'', in  {\it Perspectives on the World},
Aerts, D., Apostel, L. et all., VUBPress.

\smallskip
\noindent
Aerts, D. (1998), ``The entity and modern physics: the
creation-discovery- view of reality'', in {\it
Collection On Objects}, ed. Castelani, E., Princeton  University Press,
Princeton.

\smallskip
\noindent
Aerts, D. (1999), ``The creation discovery view and the layered
structure  of reality'', in {\it Worldviews
and the Problem of Synthesis, the Yellow Book  of Einstein meets Magritte},
ed. Aerts, D., Van Belle, H. and Van
der  Veken, J., Kluwer Academic, Dordrecht.

\smallskip
\noindent
Aerts, D. and Aerts, S. (1995-1996), ``Applications of Quantum
Statistics  in Psychological Studies of
Decision Processes'', {\it Foundations of  Science} {\bf 1}, 85.

\smallskip
\noindent
Aerts, D. and Aerts, S. (1987), ``Application of Quantum Statistics in
Psychological Studies of Decision Processes'',  in {\it Foundations of
Statistics}, eds. Van Fraassen B., Kluwer Academic, Dordrecht.

\smallskip
\noindent
Aerts, D.  Broekaert, J. Ganora L. (1999), ``The Quantum Nature of Common
Processes'',  (in preparation)

\smallskip
\noindent
Aerts, D. Coecke, B. and Smets, S. (1999), `On the Origin of Probabilities
in Quantum Mechanics: Creative and
Contextual Aspects'', in {\it Metadebates on Science, the Blue Book of Einstein
meets Magritte}, ed. Cornelis, G., Smets, S.
and Van Bendegem, J.P., Kluwer Academic, Dordrecht New York.

\smallskip
\noindent
D.  Aerts, D. and Durt, T., (1994), ``Quantum. Classical and Intermediate, an
illustrative example'', {\it Found. Phys.} {\bf 24}, 1353.

\smallskip
\noindent
Aerts, D., Durt, T. and Van Bogaert, B. (1993), ``Quantum Probability, the
Classical Limit and Non-Locality'',  in the proceedings of the {\it
International
Symposium on the Foundations of Modern Physics 1992, Helsinki, Finland}, ed. T.
Hyvonen, World Scientific, Singapore, 35.

\smallskip
\noindent
Broekaert, J. (1999), ``World Views. Elements of the Apostelian and
General Approach'', Foundations of Science, Vol.
3 ,1 , pp.

\smallskip
\noindent
CLEA (1996), Research Project: ``Integrating Worldviews: Research on the
Interdisciplinary Construction of a
Model of Reality with Ethical and Practical Relevance'' Ministry of  the
Flemish Community, dept. Science,
Innovation and Media.

\smallskip
\noindent
Grim, P. (1991), {\it The Incomplete Universe, Totality, Knowledge, and
Truth}, MIT Press, Massachusetts.

\smallskip
\noindent
Worldviews Group: Aerts, D., Apostel, L., De Moor, B., Maex, E.,
Hellemans, S., Van Belle, H., Van der
Veken, J. (1994);{\it Worldviews: From Fragmentation to Integration}, VUB
Press, Brussels

\smallskip
\noindent
Worldviews Group: Aerts, D., Apostel, L., De Moor, B., Maex, E.,
Hellemans, S., Van Belle, H., Van der
Veken, J. (1995);{\it Perspectives on the World: An Interdisciplinary
Reflection}, VUB Press, Brussels
\end{document}